\documentclass[twocolumn,aps,prl,epsfig,showpacs,superscriptaddress]{revtex4}

\usepackage{graphicx}

\begin{document}

\title{Spin-polarized tunneling spectroscopy in tunnel junctions with half-metallic electrodes}

\author{M. Bowen}
\author{A. Barth\'el\'emy}
\email{agnes.barthelemy@thalesgroup.com}
\author{M. Bibes}
\author{E. Jacquet}
\author{J.-P. Contour}
\author{A. Fert}
\affiliation{Unit\'e Mixte de Physique CNRS / Thales, Domaine de Corbeville, 91404 Orsay, France}
\author{F. Ciccacci}
\author{L. D\`uo}
\author{R. Bertacco}
\affiliation{INFM and L-NESS, Dipartimento di Fisica del Politecnico di Milano, via Anzani 52, 22100 Como,
Italy}

\date{\today}

\begin{abstract}
\vspace{0.5cm} We have studied the magnetoresistance (TMR) of tunnel junctions with electrodes of
La$_{2/3}$Sr$_{1/3}$MnO$_3$ and we show how the variation of the conductance and TMR with the bias voltage can
be exploited to obtain a precise information on the spin and energy dependence of the density of states. Our
analysis leads to a quantitative description of the band structure of La$_{2/3}$Sr$_{1/3}$MnO$_3$ and allows the
determination of the gap $\delta$ between the Fermi level and the bottom of the t$_{2g}$ minority spin band, in
good agreement with data from spin-polarized inverse photoemission experiments. This shows the potential of
magnetic tunnel junctions with half-metallic electrodes for spin-resolved spectroscopic studies.

\vspace{0.5cm}
\end{abstract}



 \maketitle

A magnetic tunnel junction (MTJ) is composed of two conducting ferromagnetic electrodes separated by a thin
insulating barrier. Its resistance depends on the relative orientation of the magnetizations of the electrodes,
a property which is called TMR (Tunneling Magnetoresistance). The TMR ratio is defined as

\begin{equation}
TMR = \frac{R_{AP}-R_P}{R_P}
\label{def_tmr}
\end{equation}

\noindent where R$_P$ and R$_{AP}$ are the junction resistances in the parallel (P) and antiparallel (AP)
configuration respectively. The MTJs are extensively investigated for the interest of the TMR in spintronic
devices such as MRAM (Magnetic Random Access Memory) or magnetic sensors \cite{wolf2002}, but they also raise
interesting fundamental problems. In this Letter, we present an example of exploitation of the TMR to obtain a
precise information on the spin and energy dependence of the density of states (DOS) of a ferromagnetic
conductor. This shows the potential of MTJs for spin-resolved spectroscopic studies.

Electron tunneling at different bias voltages (V$_{DC}$) probes different energy ranges of the DOS and this was
first used to extract information on the electronic structure of a superconducting electrode by Giaever in 1960
\cite{giaever60} : the presence of a quasi-particle gap in the DOS of the superconductor is reflected in the
voltage dependence of the current tunneling into the superconductor thus allowing a quantitative determination
of this gap. More recently, Xiang et al \cite{xiang2002} have also performed a numerical analysis of TMR vs
V$_{DC}$ curves in transition metal based MTJs to estimate the spin-dependent DOS of a Co collecting electrode.
However, the relatively narrow energy range (E$_F\pm$0.4 eV) probed by TMR is not well suited to investigate the
DOS of a wide band transition metal. We will see that the technique is more appropriate for narrow band metallic
oxides.

Also, conceptually, to extract information on the DOS above the Fermi level of a ferromagnetic collecting
electrode from the bias dependence of the TMR, it is highly desirable to use a fully spin-polarized emitting
electrode, i.e. a half-metal \cite{coey2002}. Supposing, for example, a half-metal (HM) electrode emitting only
electrons of its majority spin direction, the tunneling will probe separately the majority-spin DOS of the
collecting electrode in the P configuration (mainly at an energy eV$_{DC}$ above E$_F$) and the minority one in
the AP configuration.

The MTJs of the present work have both electrodes of the mixed-valence manganite La$_{2/3}$Sr$_{1/3}$MnO$_3$
(LSMO) which, associated with the insulating oxide SrTiO$_3$ (STO),  has unambiguously demonstrated its HM
character in tunneling experiments with a TMR ratio of 1800\% (at V$_{DC}$ = 1mV) \cite{bowen2003} and a spin
polarization (SP) of 95\%. Working with LSMO/STO/LSMO samples grown in the same pulsed laser deposition (PLD)
system and the same conditions as in Ref. \onlinecite{bowen2003} guarantees a quasi-fully spin-polarized
tunneling from the emitting electrode. Moreover, the electronic structure of the LSMO collecting electrode
displays sizeable features within the energy range accessible to TMR experiments at different bias voltages. In
particular we will see that the gap $\delta$ separating the Fermi level from the bottom of the minority spin
t$_{2g}$ conduction band of LSMO is in the range of a few tenths of eV and well accessible in TMR. The
LSMO/STO/LSMO MTJs are thus ideally suited for spin-resolved spectroscopic investigations. As quantitatively
predicted by Bratkovsky \cite{bratkovsky97}, a very large TMR is expected at low bias, and the TMR(V$_{DC}$)
dependence should present a sharp decrease when V$_{DC}$ becomes equal to $\delta$/e due to the opening of new
conducting channels in the antiparallel configuration. From the bias dependence of the TMR we will extract the
gap $\delta$ between the Fermi level (E$_F$) and the bottom of the minority spin t$_{2g}$ band, which we find to
amount at 340$\pm$20 meV. To confirm our observation, we have performed spin-polarized inverse photoemission
(SPIPE) on a LSMO/STO interface to explore the DOS of LSMO above E$_F$ in an alternative spin-resolved
technique. A quite good agreement is found between the two values of $\delta$.

\begin{figure}[!h]
 \includegraphics[keepaspectratio=true,width=\columnwidth]{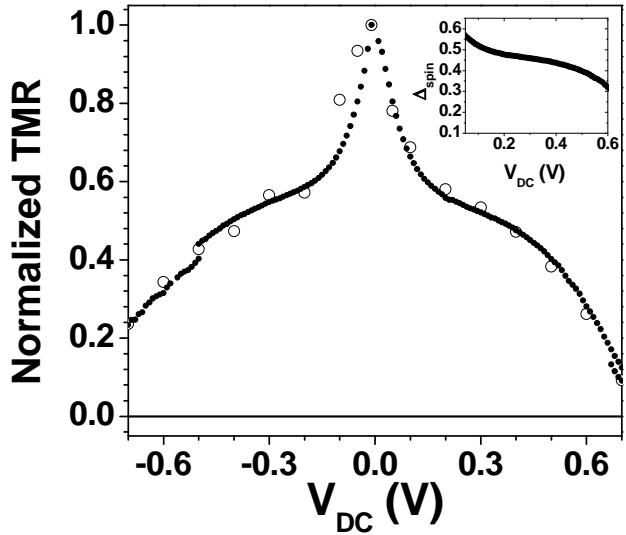}
 \caption{Bias dependence of TMR (normalized by the value found at V$_{DC}$ = -10mV) at T=4K for a LSMO/STO/LSMO junction from R(H) (open circles) and P/AP I(V$_{DC}$) data (filled circles).
 Insert: spin asymmetry $\Delta_{\text{spin}}=(I_{P}-I_{AP})/(I_{P}+I_{AP})=P^{2}$.}
 \label{tmr(v)}
\end{figure}

The heterostructures of this study have been grown by pulsed laser deposition (PLD) and are fully epitaxial
\cite{lyonnetLSMO}. A detailed structural and spectroscopic characterization of the STO/LSMO(001) interface by
High Resolution Transmission Electron Microscopy and Electron Energy Loss Spectroscopy has already been
published \cite{pailloux2002}. Junctions as small as a few microns square were defined by standard optical
lithography \cite{bowen2003}. Since the polarization at the interface is the relevant spin polarization in
tunneling measurements \cite{yamada2004}, SPIPE experiments \cite{chiaiaSPIPE93} were carried out on an
epitaxial LSMO film capped by a thin (0.8 nm) layer of STO. The small probe depth of this technique ($\sim$ 10
$\rm{\AA}$) allows to probe the unoccupied part of the manganite DOS at the interface with STO. Furthermore the
STO overlayer acts as a capping layer for LSMO, so that problems arising from surface contamination in SPIPE
experiments are reduced with respect to the case of the LSMO free surface \cite{bertaccoLSMO2002}.

In Fig.\ref{tmr(v)} we present the bias dependence of TMR for a 2x6 mm$^2$ LSMO/STO/LSMO magnetic tunnel
junction measured at T=4K.  Three different regimes can be distinguished. At low bias, the TMR amplitude drops
rapidly with bias before levelling off at about ±$\pm$120mV. For clarity the data have been normalized to the
350\% value found at V$_{DC}$ = -10mV. In an intermediate bias range, 120 mV $\leq |$V$_{DC}|\leq$ 340 mV, the
TMR decreases only very slowly. Finally, in the high bias regime for $|$V$_{DC}|\geq$ 0.34 V, the TMR decreases
rapidly again. The small asymmetry between negative and positive bias, especially noticeable in this high bias
regime, probably reflects a slight difference in the chemical structure of the upper and lower LSMO/STO
interfaces in our junctions as observed by Electron Energy Loss Spectroscopy \cite{sametEELS2003EPJB}.

\begin{figure}[!h]
\includegraphics[keepaspectratio=true,width=\columnwidth]{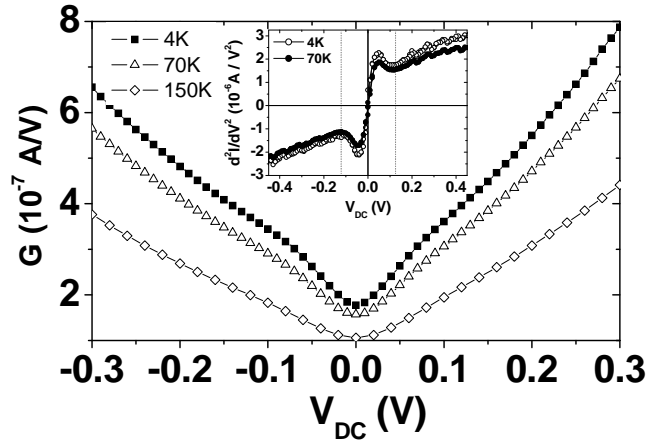}
\caption{Temperature evolution of the parallel conductance from
 4K (closed squares) to higher temperatures (open symbols).
 Insert: bias dependence of the parallel conductance derivative at 4K and 70K.}
\label{zba}
\end{figure}

The drop in TMR observed at low bias at T=4K is associated with a zero-bias conductance anomaly
\cite{gu2001,zhang98,moodera98} which is occurring also for $|$V$_{DC}|\leq$ 120 mV, as illustrated in
Fig.\ref{zba}. Both the conductance anomaly and the drop of TMR at low bias are what is expected \cite{gu2001}
from spin wave excitations induced by the tunneling electrons at the LSMO/STO interface. As expected for spin
wave excitations, the zero-bias anomaly progressively disappears as temperature increases, see Fig.\ref{zba}. A
more detailed report on this zero bias anomaly will be published elsewhere. Here we conclude that the low bias
regime, with a change of TMR with bias predominantly due to spin wave excitations, is not appropriate to study
DOS effects. On the other hand, these variations clearly saturate at a bias of about 120mV independently of the
temperature as can be seen, for example, in the insert of Fig.\ref{zba} which shows the derivative of the
conductance as a function of the bias at several temperatures.

\begin{figure}[!h]
 \includegraphics[keepaspectratio=true,width=\columnwidth]{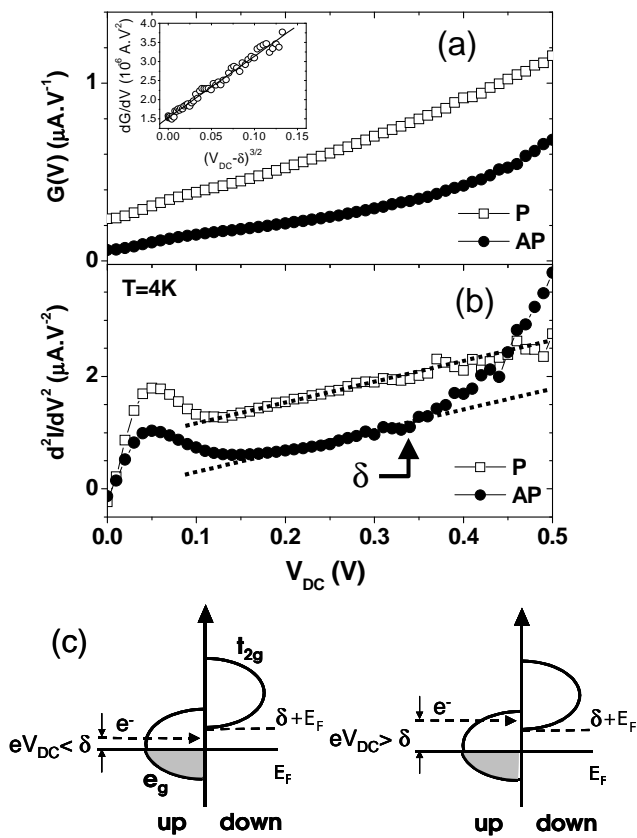}
 \caption{(a) Bias dependence of the parallel and antiparallel conductances G$_P$=(dI/dV)$_P$
 and G$_{AP}$=(dI/dV)$_{AP}$ at T=4K. (b) Bias dependence of the conductance derivatives showing the onset
 of tunneling into the minority-spin t$_{2g}$ sub-band for the antiparallel conductance
 at 350mV. Dotted lines are guides to the eye. The conductance derivative in the AP configuration is plotted vs (V$_{DC}$-$\delta$)$^{3/2}$
 in the inset of (a) ; symbols are experimental data and the line is a linear fit. (c)
 Schematic representation of tunneling towards a half-metallic collecting electrode for
 eV$_{DC}$ smaller (left panel) and larger (right panel) than $\delta$.}
 \label{g(v)}
\end{figure}

As shown in figure \ref{tmr(v)}, the TMR decreases slowly as a function of the bias in the intermediate regime,
120 mV $\leq |$V$_{DC}|\leq$ 340 mV. The insert of figure 1 confirms the fairly constant evolution of the spin
asymmetry $\Delta_{spin}$ (defined as $\Delta_{spin}$=(I$_P$-I$_{AP}$)/(I$_P$+I$_{AP}$)=P$^2$) in this bias
range. This intermediate regime of slow variation is followed by a much more rapid decrease of the TMR after an
inflection point at $|$V$_{DC}|$=0.34 V. Looking separately at the conductances in the parallel (P) and
antiparallel (AP) states on Fig.\ref{g(v)}a, we see that the conductance in the P state, (dI/dV)$_P$, which
reflects the tunneling between the majority spin states of the two electrodes, increases smoothly within the
bias range 120 mV $\leq |$V$_{DC}|\leq$ 500 mV. In contrast, the conductance in the AP configuration,
(dI/dV)$_{AP}$, which reflects the tunneling from majority spin states in the emitting electrode to the minority
spin band in the collecting one, shows an upturn around $|$V$_{DC}|$=0.34 V. This change of behavior is even
clearer on the derivative of the conductance in the AP state (reported on fig.\ref{g(v)}b). This value of
$|$V$_{DC}|$=340 mV, has been corroborated for both interfaces of several MTJs to within an error of 40 meV.

The inflection point at 340mV in the TMR(V$_{DC}$) curve together with the slope change in the variation of
d$^2$I/dV$^2$ with V$_{DC}$ in the AP state indicates a strong increase of the DOS of the minority spin sub-band
at about this energy. Such features are consistent with the band structure of LSMO proposed in fig \ref{g(v)}.c
with the bottom of a minority t$_{2g}$ sub-band located at $\delta$=(340$\pm$20) meV above the Fermi level.
Increasing the bias voltage above 340 mV in the AP configuration opens a tunneling channel between the majority
spin band of the emitting electrode to the minority spin band of the collecting one, which gives rise to the
upturn of the conductance of the AP state and to the drop of the TMR.

Beyond a picture in terms of the only DOS, we have to consider that the symmetry matching of the wave functions
is also an important factor of the tunneling probability \cite{butler2001,yuasa2004}. In the P configuration of
our junctions, according to the calculation of the complex band structure of STO by Bellini \cite{bellini}, the
majority spin e$_g$ states of the LSMO electrodes are predominantly connected by a slowly decaying MIGS (Metal
Induced Gap State) of symmetry $\Delta_1$. On the contrary, for the tunneling between the majority spin e$_g$
and the minority spin t$_{2g}$ band in the AP configuration at large bias, such a connection by a slowly
decaying MIGS does not exist. Symmetry breaking by the disorder (interface roughness, disordered distribution of
La and Sr, which are not taken into account in the calculation) can however introduce some coupling but less
efficiently than with the $\Delta_1$ MIGS of the parallel configuration. This possibly explains that, in spite
of the similar amplitude of the DOS in the majority spin e$_g$ and minority spin t$_{2g}$ bands
\cite{banach2004}, the conductance G$_{AP}$ of the AP configuration above 0.34 eV does not catch rapidly G$_P$
and remains markedly smaller (60\% of G$_P$ at 0.5 eV). The relatively gradual increase of G$_{AP}$ above 0.34
eV can also be due to some broadening of the t$_{2g}$ band edge by interface disorder.

Discarding the zero-bias anomaly, the variation of the TMR and the conductance with V$_{DC}$ can actually be
compared to the DOS-based predictions of Bratkovsky \cite{bratkovsky97} for HM/I/HM MTJs. As in our
interpretation, Bratkovsky predicts an increase of the conductance in the AP state associated with a drop of the
TMR above the threshold bias corresponding to the minority gap ($\delta$=0.3 eV in ref \cite{bratkovsky97} and
0.34 eV in our case). Furthermore, above this threshold value, a variation of the conductance in the AP state as
G$_{AP}$=(V$_{DC}$-$\delta$)$^{5/2}$ , corresponding to a variation of the derivative of the conductance as
dG$_{AP}$/dV=(V$_{DC}$-$\delta$)$^{3/2}$ is expected. As shown in the insert of Fig.\ref{g(v)}a, which presents
the experimental variation of the derivative of the conductance in the antiparallel state as a function of
(V-$\delta$)$^{3/2}$ with $\delta$=0.34 eV, a good agreement is found between the theoretical prediction and our
experimental results.

\begin{figure}[!h]
 \includegraphics[keepaspectratio=true,width=0.9\columnwidth]{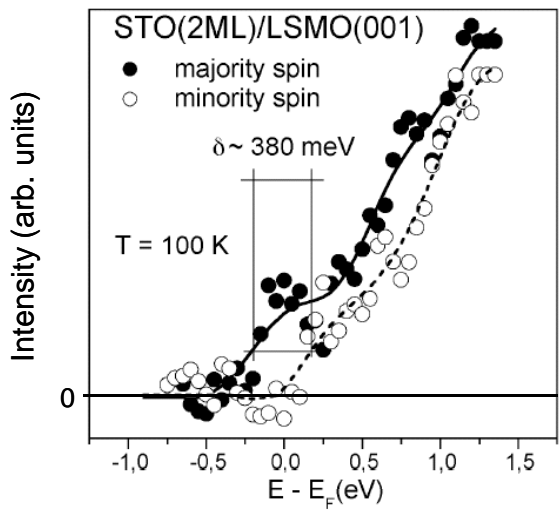}
 \caption{Spin-resolved spectra taken at 100 K for a STO/LSMO interface. A smoothing of experimental data at 100
 K is plotted for the two spin channels.} \label{spipe}
\end{figure}

We have performed spin polarized inverse photoemission (SPIPE) experiments onto a LSMO layer covered by two unit
cells of STO to confirm the experimental value of $\delta$ and to see if the band structure is probed in a
similar way by spin-dependent tunneling and SPIPE. The SPIPE spectra taken at 100K in an energy range close to
E$_F$ for a LSMO/STO bilayer are presented in Fig. \ref{spipe}. Two distinct line-shapes for the majority- (full
dots) and minority-spin channels (empty dots) can be clearly distinguished. A sizable signal is visible at E$_F$
in the majority-spin channel \cite{comment-spipe} but appears only at an higher energy for the minority-spin.
This clearly indicates that the sample is metallic for majority electrons and insulating for minority electrons.
Due to both the very low counting rate at E$_F$ and the rescaling procedure to 100\% polarization of the
incident electron beam \cite{chiaiaSPIPE93}, the data present a significant scattering. This prevents a precise
determination of the spin polarization at E$_F$, which nevertheless can be estimated to be $\sim$ 90\%. The
t$_{2g}$ band responsible for the delayed onset of the minority signal in the SPIPE spectrum thus defines a gap
$\delta$ between E$_F$ and the low-energy edge of the minority-spin sub-band. The extent of this minority gap
can be estimated from the energy difference between the minority- and majority-channel onsets, as its absolute
position on the energy scale is affected by experimental broadening. We obtain a value of 380$\pm$50 meV for
$\delta$, as shown in Fig.\ref{spipe}, where the result of smoothing the experimental data, coherently with our
energy resolution, is also plotted. This value is close to what is obtained for a free LSMO surface
\cite{bertaccoLSMO2002} and in good agreement with what we find from spin-dependent tunneling.

The values of $\delta$ derived from theoretical calculations for bulk LSMO range from 0 \cite{livesay99} to 1.6
eV \cite{banach2004}, a broad range including our experimental value. We however note that the value of $\delta$
at the interface with STO can be different from that calculated for bulk LSMO, due to bandwidth contraction and
possible Anderson localization effects \cite{calderon99}. The importance of this type of effect could be
addressed by systematic studies of different types of interfaces (LSMO/TiO$_2$, LSMO/LaAlO$_3$).

In conclusion, the interpretation of the bias dependence of the conductance and TMR of LSMO/STO/LSMO MTJs
provides us with a quantitative spin-resolved information on the band structure at the LSMO/STO interface and,
especially, on the gap between the Fermi energy of LSMO and the bottom of the t$_{2g}$ minority spin band. We
have also found that this band structure is quite consistent with spin-polarized inverse photoemission
measurements. The agreement between these two types of experiments strengthens the interpretation of the TMR on
the basis of the Bratkovsky model for tunneling between two half-metals. It also turns out that a half-metallic
emitting electrode confers to MTJ a potential for spin-resolved spectroscopy capability which can be of high
interest to probe the electronic structure of some new ferromagnetic materials, diluted magnetic semiconductors
for example.

\acknowledgments{We would like to thank Mike Coey, Wulf Wulfhekel, Daniel Lacour, Patrick LeClair, Walter
Temmerman, Grzgorz Banach and Pierre Seneor for stimulating discussions. This work was financed in part by the
AMORE European contract (G5RD-CT-2000-00138) and the "Computational Magnetoelectronics" Research and Training
Network.}


\vspace{0.5em}


\begin{thebibliography}{17}
\expandafter\ifx\csname natexlab\endcsname\relax\def\natexlab#1{#1}\fi \expandafter\ifx\csname
bibnamefont\endcsname\relax
  \def\bibnamefont#1{#1}\fi
\expandafter\ifx\csname bibfnamefont\endcsname\relax
  \def\bibfnamefont#1{#1}\fi
\expandafter\ifx\csname citenamefont\endcsname\relax
  \def\citenamefont#1{#1}\fi
\expandafter\ifx\csname url\endcsname\relax
  \def\url#1{\texttt{#1}}\fi
\expandafter\ifx\csname urlprefix\endcsname\relax\def\urlprefix{URL }\fi \providecommand{\bibinfo}[2]{#2}
\providecommand{\eprint}[2][]{\url{#2}}


\bibitem[{\citenamefont{Wolf et~al.}(2002)}]{wolf2002}
\bibinfo{author}{\bibfnamefont{S.A.}~\bibnamefont{Wolf}},
\bibinfo{author}{\bibfnamefont{D.D}~\bibnamefont{Awschalom}},
\bibinfo{author}{\bibfnamefont{R.A.}~\bibnamefont{Buhrman}},
\bibinfo{author}{\bibfnamefont{J.M.}~\bibnamefont{Daughton}},
\bibinfo{author}{\bibfnamefont{S.}~\bibnamefont{von Moln\`ar}},
\bibinfo{author}{\bibfnamefont{M.L.}~\bibnamefont{Roukes}},
\bibinfo{author}{\bibfnamefont{A.Y.}~\bibnamefont{Chtchelkanova}} \bibnamefont{and}
\bibinfo{author}{\bibfnamefont{D.M.}~\bibnamefont{Treger}},
  \bibinfo{journal}{Science} \textbf{\bibinfo{volume}{294}},
  \bibinfo{pages}{1488} (\bibinfo{year}{2002}).


\bibitem[{\citenamefont{Giaever}(1960)}]{giaever60}
\bibinfo{author}{\bibfnamefont{I.}~\bibnamefont{Giaever}},
  \bibinfo{journal}{Phys. Rev. Lett.} \textbf{\bibinfo{volume}{5}},
  \bibinfo{pages}{147} (\bibinfo{year}{1960}).


\bibitem[{\citenamefont{Xiang et~al.}(2002)\citenamefont{Xiang, Zhu, Du,
  Landry, and Xiao}}]{xiang2002}
\bibinfo{author}{\bibfnamefont{X.H.}~\bibnamefont{Xiang}},
  \bibinfo{author}{\bibfnamefont{T.}~\bibnamefont{Zhu}},
  \bibinfo{author}{\bibfnamefont{J.}~\bibnamefont{Du}},
  \bibinfo{author}{\bibfnamefont{G.}~\bibnamefont{Landry}}, \bibnamefont{and}
  \bibinfo{author}{\bibfnamefont{J.Q.}~\bibnamefont{Xiao}},
  \bibinfo{journal}{Phys. Rev. B} \textbf{\bibinfo{volume}{66}},
  \bibinfo{pages}{174407} (\bibinfo{year}{2002}).


\bibitem[{\citenamefont{Coey and Venkatesan}(2002)}]{coey2002}
\bibinfo{author}{\bibfnamefont{J.M.D.}~\bibnamefont{Coey}} \bibnamefont{and}
  \bibinfo{author}{\bibfnamefont{M.}~\bibnamefont{Venkatesan}},
  \bibinfo{journal}{J. Appl. Phys.} \textbf{\bibinfo{volume}{91}},
  \bibinfo{pages}{8345} (\bibinfo{year}{2002}).


\bibitem[{\citenamefont{Bowen et~al.}(2003)}]{bowen2003}
\bibinfo{author}{\bibfnamefont{M.}~\bibnamefont{Bowen}},
\bibinfo{author}{\bibfnamefont{M.}~\bibnamefont{Bibes}},
\bibinfo{author}{\bibfnamefont{A.}~\bibnamefont{Barth\'el\'emy}},
\bibinfo{author}{\bibfnamefont{J.-P.}~\bibnamefont{Contour}},
\bibinfo{author}{\bibfnamefont{A.}~\bibnamefont{Anane}},
\bibinfo{author}{\bibfnamefont{Y.}~\bibnamefont{Lema\^itre}} \bibnamefont{and}
\bibinfo{author}{\bibfnamefont{A.}~\bibnamefont{Fert}},
  \bibinfo{journal}{Appl. Phys. Lett.} \textbf{\bibinfo{volume}{82}},
  \bibinfo{pages}{233} (\bibinfo{year}{2003}).


\bibitem[{\citenamefont{Bratkovsky}(1997)}]{bratkovsky97}
\bibinfo{author}{\bibfnamefont{A.M.}~\bibnamefont{Bratkovsky}},
  \bibinfo{journal}{Phys. Rev. B} \textbf{\bibinfo{volume}{56}},
  \bibinfo{pages}{2344} (\bibinfo{year}{1997}).


\bibitem[{\citenamefont{Lyonnet et~al.}(2000)\citenamefont{Lyonnet, Maurice,
  Hytch, Michel, and Contour}}]{lyonnetLSMO}
\bibinfo{author}{\bibfnamefont{R.}~\bibnamefont{Lyonnet}},
  \bibinfo{author}{\bibfnamefont{J.-L.} \bibnamefont{Maurice}},
  \bibinfo{author}{\bibfnamefont{M.}~\bibnamefont{Hytch}},
  \bibinfo{author}{\bibfnamefont{D.}~\bibnamefont{Michel}}, \bibnamefont{and}
  \bibinfo{author}{\bibfnamefont{J.-P.} \bibnamefont{Contour}},
  \bibinfo{journal}{Appl. Surf. Sci.} \textbf{\bibinfo{volume}{162-163}},
  \bibinfo{pages}{245} (\bibinfo{year}{2000}).


\bibitem[{\citenamefont{Pailloux et~al.}(2002)}]{pailloux2002}
\bibinfo{author}{\bibfnamefont{F.}~\bibnamefont{Pailloux}},
\bibinfo{author}{\bibfnamefont{D.}~\bibnamefont{Imhoff}},
\bibinfo{author}{\bibfnamefont{T.}~\bibnamefont{Sikora}},
\bibinfo{author}{\bibfnamefont{A.}~\bibnamefont{Barth\'el\'emy}},
\bibinfo{author}{\bibfnamefont{J.-L.}~\bibnamefont{Maurice}},
\bibinfo{author}{\bibfnamefont{J.-P.}~\bibnamefont{Contour}},
\bibinfo{author}{\bibfnamefont{C.}~\bibnamefont{Colliex}} \bibnamefont{and}
\bibinfo{author}{\bibfnamefont{A.}~\bibnamefont{Fert}},
\bibinfo{journal}{Phys. Rev. B}
  \textbf{\bibinfo{volume}{66}}, \bibinfo{pages}{014417}
  (\bibinfo{year}{2002}).


\bibitem[{\citenamefont{Yamada et~al.}(2004)}]{yamada2004}
\bibinfo{author}{\bibfnamefont{H.}~\bibnamefont{Yamada}},
\bibinfo{author}{\bibfnamefont{Y.}~\bibnamefont{Ogawa}},
\bibinfo{author}{\bibfnamefont{Y.}~\bibnamefont{Ishii}},
\bibinfo{author}{\bibfnamefont{H.}~\bibnamefont{Sato}},
\bibinfo{author}{\bibfnamefont{M.}~\bibnamefont{Kawasaki}},
\bibinfo{author}{\bibfnamefont{H.}~\bibnamefont{Akoh}},  \bibnamefont{and}
\bibinfo{author}{\bibfnamefont{Y.}~\bibnamefont{Tokura}},
\bibinfo{journal}{Science}
  \textbf{\bibinfo{volume}{305}}, \bibinfo{pages}{646}
  (\bibinfo{year}{2004}).


\bibitem[{\citenamefont{G.Chiaia et~al.}(1993)\citenamefont{G.Chiaia, Rossi,
  Mazzolari, and Ciccacci}}]{chiaiaSPIPE93}
\bibinfo{author}{\bibnamefont{G.Chiaia}}, \bibinfo{author}{\bibfnamefont{S.}
  \bibnamefont{De Rossi}},
  \bibinfo{author}{\bibfnamefont{L.}~\bibnamefont{Mazzolari}},
  \bibnamefont{and} \bibinfo{author}{\bibfnamefont{F.}~\bibnamefont{Ciccacci}},
  \bibinfo{journal}{Phys. Rev. B} \textbf{\bibinfo{volume}{48}},
  \bibinfo{pages}{11298} (\bibinfo{year}{1993}).


\bibitem[{\citenamefont{Bertacco et~al.}(2002)}]{bertaccoLSMO2002}
\bibinfo{author}{\bibfnamefont{R.}~\bibnamefont{Bertacco}}
  \bibnamefont{et~al.}, \bibinfo{journal}{J. Magn. Magn. Mater.}
  \textbf{\bibinfo{volume}{242-245}}, \bibinfo{pages}{710}
  (\bibinfo{year}{2002}).


\bibitem[{\citenamefont{Samet et~al.}(2003)}]{sametEELS2003EPJB}
\bibinfo{author}{\bibfnamefont{L.}~\bibnamefont{Samet}} \bibnamefont{et~al.},
  \bibinfo{journal}{unpublished}  (\bibinfo{year}{2003}).


\bibitem[{\citenamefont{Moodera et~al.}(1998)}]{moodera98}
\bibinfo{author}{\bibfnamefont{J.S.}~\bibnamefont{Moodera}},
  \bibinfo{author}{\bibfnamefont{J.}~\bibnamefont{Nowak}}, \bibnamefont{and}
  \bibinfo{author}{\bibfnamefont{R.J.M.}~\bibnamefont{vandeVeerdonk}},
  \bibinfo{journal}{Phys. Rev. Lett.} \textbf{\bibinfo{volume}{80}},
  \bibinfo{pages}{2941} (\bibinfo{year}{1998}).


\bibitem[{\citenamefont{Gu et~al.}(2001)\citenamefont{Gu, Sheng, and
  Ting}}]{gu2001}
\bibinfo{author}{\bibfnamefont{R.Y.}~\bibnamefont{Gu}},
  \bibinfo{author}{\bibfnamefont{L.}~\bibnamefont{Sheng}}, \bibnamefont{and}
  \bibinfo{author}{\bibfnamefont{C.S.}~\bibnamefont{Ting}},
  \bibinfo{journal}{Phys. Rev. B} \textbf{\bibinfo{volume}{63}},
  \bibinfo{pages}{220406(R)} (\bibinfo{year}{2001}).


\bibitem[{\citenamefont{Zhang et~al.}(1998)\citenamefont{Zhang, Levy, Marley,
  and Parkin}}]{zhang98}
\bibinfo{author}{\bibfnamefont{S.}~\bibnamefont{Zhang}},
  \bibinfo{author}{\bibfnamefont{P.~M.} \bibnamefont{Levy}},
  \bibinfo{author}{\bibfnamefont{A.~C.} \bibnamefont{Marley}},
  \bibnamefont{and} \bibinfo{author}{\bibfnamefont{S.~S.~P.}
  \bibnamefont{Parkin}}, \bibinfo{journal}{Phys. Rev. Lett.}
  \textbf{\bibinfo{volume}{79}}, \bibinfo{pages}{3744} (\bibinfo{year}{1998}).


\bibitem[{\citenamefont{Butler et~al.}(2001)}]{butler2001}
\bibinfo{author}{\bibfnamefont{W.H.}~\bibnamefont{Butler}},
\bibinfo{author}{\bibfnamefont{X.-G.}~\bibnamefont{Zhang}},
\bibinfo{author}{\bibfnamefont{T.C.}~\bibnamefont{Schulthess}}, \bibnamefont{and}
\bibinfo{author}{\bibfnamefont{J.M.}~\bibnamefont{MacLaren}},
  \bibinfo{journal}{Phys. Rev. B} \textbf{\bibinfo{volume}{63}},
  \bibinfo{pages}{054416}
  (\bibinfo{year}{2001}).


\bibitem[{\citenamefont{Yuasa et~al.}(2004)}]{yuasa2004}
\bibinfo{author}{\bibfnamefont{S.}~\bibnamefont{Yuasa}},
\bibinfo{author}{\bibfnamefont{T.}~\bibnamefont{Nagahama}},
\bibinfo{author}{\bibfnamefont{A.}~\bibnamefont{Fukushima}},
\bibinfo{author}{\bibfnamefont{Y.}~\bibnamefont{Suzuki}}, \bibnamefont{and}
  \bibinfo{author}{\bibfnamefont{K.}~\bibnamefont{Ando}},
  \bibinfo{journal}{Nature Materials advance online publication},
  \bibinfo{pages}{doi:10.1038/nmat1257}.


\bibitem[{\citenamefont{V. Bellini}(2000)}]{bellini}
\bibinfo{author}{\bibfnamefont{V.}~\bibnamefont{Bellini}},
  \bibinfo{journal}{Electronic structure of low-dimensional magnetic systems, Ph.D thesis, Rheinisch-Westfälischen Technischen Hochschule, Aachen, Germany}
  (\bibinfo{year}{2000}).


\bibitem[{\citenamefont{Banach et~al.}(2004)\citenamefont{Banach, and
  Temmerman}}]{banach2004}
\bibinfo{author}{\bibfnamefont{G.}~\bibnamefont{Banach}}, \bibnamefont{and}
  \bibinfo{author}{\bibfnamefont{W.M.}~\bibnamefont{Temmerman}},
  \bibinfo{journal}{Phys. Rev. B} \textbf{\bibinfo{volume}{69}},
  \bibinfo{pages}{054427}
  (\bibinfo{year}{2004}).


\bibitem[{\citenamefont{comment-spipe}(2000)}]{comment-spipe}
\bibinfo{journal}{Due to the finite resolution of the SPIPE experiment, a non vanishing signal is actually seen even for E$<$E$_F$.
The overall resolution, including the electron beam dispersion and the band pass full width at half-maximum of
the photon detector (by far the dominating term), is 0.7 eV, see F. Ciccacci, S. De Rossi, A. Taglia, S.
Crampin, J. Phys.: Cond. Mat. 6, 7227 (1994)}.


\bibitem[{\citenamefont{Livesay et~al.}(1999)\citenamefont{Livesay, West,
  Dugdale, Santi, and Jarlborg}}]{livesay99}
\bibinfo{author}{\bibfnamefont{E.}~\bibnamefont{Livesay}},
  \bibinfo{author}{\bibfnamefont{R.}~\bibnamefont{West}},
  \bibinfo{author}{\bibfnamefont{S.}~\bibnamefont{Dugdale}},
  \bibinfo{author}{\bibfnamefont{G.}~\bibnamefont{Santi}}, \bibnamefont{and}
  \bibinfo{author}{\bibfnamefont{T.}~\bibnamefont{Jarlborg}},
  \bibinfo{journal}{J. Phys.: Cond. Mat.} \textbf{\bibinfo{volume}{11}},
  \bibinfo{pages}{L279} (\bibinfo{year}{1999}).


\bibitem[{\citenamefont{Calder\'on et~al.}(1999)\citenamefont{Calder\'on, Brey,
  and Guinea}}]{calderon99}
\bibinfo{author}{\bibfnamefont{M.J.}~\bibnamefont{Calder\'on}},
  \bibinfo{author}{\bibfnamefont{L.}~\bibnamefont{Brey}}, \bibnamefont{and}
  \bibinfo{author}{\bibfnamefont{F.}~\bibnamefont{Guinea}},
  \bibinfo{journal}{Phys. Rev. B} \textbf{\bibinfo{volume}{60}},
  \bibinfo{pages}{6698} (\bibinfo{year}{1999}).


\end{thebibliography}
\end{document}